\documentclass[
aps,prl,reprint,superscriptaddress,
]{revtex4-2}

\usepackage{graphicx}
\usepackage{dcolumn}
\usepackage{bm}
\usepackage{etoolbox}
\usepackage{color}
\usepackage{xcolor}
\usepackage{amsmath} 
\usepackage{enumitem}
\usepackage{float}

\begin{document}


\title{Laser-intensity-spike-dominated hot electron generation from two-plasmon decay instability driven by moderate-bandwidth pulses
}
\author{C. Yao}
\thanks{These authors contributed equally to this work}
\affiliation{ 
Department of Plasma Physics and Fusion Engineering and CAS Key Laboratory of Frontier Physics in Controlled Nuclear Fusion, University of Science and Technology of China, Hefei, Anhui 230026, China\looseness=-1
}%
\affiliation{ 
Institute of Applied Physics and Computational Mathematics, Beijing 100088, China
}%

\author{Z. H. Cai}%
\thanks{These authors contributed equally to this work}
\affiliation{ 
Department of Plasma Physics and Fusion Engineering and CAS Key Laboratory of Frontier Physics in Controlled Nuclear Fusion, University of Science and Technology of China, Hefei, Anhui 230026, China\looseness=-1
}%
\author{X. Wang}%
\thanks{These authors contributed equally to this work}
\affiliation{ 
Shanghai Institute of Laser Plasma, China Academy of Engineering Physics, Shanghai 201899, China 
}%

\author{X. C. Wang}%
\affiliation{ 
Department of Plasma Physics and Fusion Engineering and CAS Key Laboratory of Frontier Physics in Controlled Nuclear Fusion, University of Science and Technology of China, Hefei, Anhui 230026, China\looseness=-1
}%

\author{H. R. Yin}%
\affiliation{ 
Department of Plasma Physics and Fusion Engineering and CAS Key Laboratory of Frontier Physics in Controlled Nuclear Fusion, University of Science and Technology of China, Hefei, Anhui 230026, China\looseness=-1
}%

\author{Z. A. Zhu}%
\affiliation{ 
Department of Plasma Physics and Fusion Engineering and CAS Key Laboratory of Frontier Physics in Controlled Nuclear Fusion, University of Science and Technology of China, Hefei, Anhui 230026, China\looseness=-1
}%

\author{C. W. Lian}%
\affiliation{ 
State Key Laboratory of High Temperature Gas Dynamics, School of Engineering Science, University of Science and Technology of China, Hefei 230026, China
}%

\author{Y. Ji}%
\affiliation{ 
State Key Laboratory of High Temperature Gas Dynamics, School of Engineering Science, University of Science and Technology of China, Hefei 230026, China
}%

\author{X. Jiang}%
\affiliation{ 
State Key Laboratory of High Temperature Gas Dynamics, School of Engineering Science, University of Science and Technology of China, Hefei 230026, China
}%

\author{S. M. Xu}%
\affiliation{ 
State Key Laboratory of High Temperature Gas Dynamics, School of Engineering Science, University of Science and Technology of China, Hefei 230026, China
}%

\author{Y. Y. Yao}%
\affiliation{ 
State Key Laboratory of High Temperature Gas Dynamics, School of Engineering Science, University of Science and Technology of China, Hefei 230026, China
}%

\author{L. Y. Yang}%
\affiliation{ 
Shanghai Institute of Laser Plasma, China Academy of Engineering Physics, Shanghai 201899, China 
}%

\author{J. N. Zhang}%
\affiliation{ 
Shanghai Institute of Laser Plasma, China Academy of Engineering Physics, Shanghai 201899, China 
}%

\author{D. Meng}%
\affiliation{ 
School of Physics and Electronics, Hunan University, Changsha 410082, China
}%

\author{T. Peng}%
\affiliation{ 
School of Physics and Electronics, Hunan University, Changsha 410082, China
}%

\author{H. Wen}%
\affiliation{ 
School of Physics and Electronics, Hunan University, Changsha 410082, China
}%

\author{C. Z. Xiao}%
\affiliation{ 
School of Physics and Electronics, Hunan University, Changsha 410082, China
}%

\author{K. Y. Meng}%
\affiliation{ 
Department of Plasma Physics and Fusion Engineering and CAS Key Laboratory of Frontier Physics in Controlled Nuclear Fusion, University of Science and Technology of China, Hefei, Anhui 230026, China\looseness=-1
}%

\author{J. Li}%
\thanks{Corresponding author: junlisu@ustc.edu.cn}
\affiliation{ 
Department of Plasma Physics and Fusion Engineering and CAS Key Laboratory of Frontier Physics in Controlled Nuclear Fusion, University of Science and Technology of China, Hefei, Anhui 230026, China\looseness=-1
}%
\affiliation{ 
Collaborative Innovation Center of IFSA (CICIFSA), Shanghai Jiao Tong University, Shanghai 200240, China
}%

\author{R. Yan}%
\thanks{Corresponding author: ruiyan@ustc.edu.cn}
\affiliation{ 
State Key Laboratory of High Temperature Gas Dynamics, School of Engineering Science, University of Science and Technology of China, Hefei 230026, China
}%
\affiliation{ 
Collaborative Innovation Center of IFSA (CICIFSA), Shanghai Jiao Tong University, Shanghai 200240, China
}%

\author{P. Yuan}%
\thanks{Corresponding author: yuan.peng@sjtu.edu.cn}
\affiliation{ 
Shanghai Tsung-Dao Lee Institute, State Key Laboratory of Dark Matter Physics, Shanghai Jiao Tong University, Shanghai 201210, China 
}%

\author{Z. Zhang}%
\affiliation{ 
Beijing National Laboratory for Condensed Matter Physics, Institute of Physics, Chinese Academy of Sciences, Beijing 100190, China 
}%

\author{L. Hao}
\affiliation{ 
Institute of Applied Physics and Computational Mathematics, Beijing 100088, China
}%

\author{Q. Jia}%
\affiliation{ 
Department of Plasma Physics and Fusion Engineering and CAS Key Laboratory of Frontier Physics in Controlled Nuclear Fusion, University of Science and Technology of China, Hefei, Anhui 230026, China\looseness=-1
}%

\author{W. Feng}%
\affiliation{ 
Shanghai Institute of Laser Plasma, China Academy of Engineering Physics, Shanghai 201899, China 
}%

\author{H. H. An}%
\affiliation{ 
Shanghai Institute of Laser Plasma, China Academy of Engineering Physics, Shanghai 201899, China 
}%

\author{H. Y. Liu}%
\affiliation{Key Laboratory of High Power Laser and Physics, Shanghai Institute of Optics and Fine Mechanics, Chinese Academy of Sciences, Shanghai 201800, China\looseness=-1
}%

\author{Z. Y. Xie}%
\affiliation{ 
Shanghai Institute of Laser Plasma, China Academy of Engineering Physics, Shanghai 201899, China 
}%

\author{P. P. Wang}%
\affiliation{ 
Shanghai Institute of Laser Plasma, China Academy of Engineering Physics, Shanghai 201899, China 
}%

\author{C. Wang}%
\affiliation{ 
Shanghai Institute of Laser Plasma, China Academy of Engineering Physics, Shanghai 201899, China 
}%

\author{A. Lei}%
\affiliation{ 
Shanghai Institute of Laser Plasma, China Academy of Engineering Physics, Shanghai 201899, China 
}%

\author{X. H. Zhao}%
\affiliation{ 
Shanghai Institute of Laser Plasma, China Academy of Engineering Physics, Shanghai 201899, China 
}%

\author{Z. H. Fang}%
\affiliation{ 
Shanghai Institute of Laser Plasma, China Academy of Engineering Physics, Shanghai 201899, China 
}%

\author{W. Wang}%
\thanks{Corresponding author: wei\_wang@fudan.edu.cn}
\affiliation{ 
Shanghai Institute of Laser Plasma, China Academy of Engineering Physics, Shanghai 201899, China 
}%

\author{Y. Q. Gu}%
\affiliation{ 
Shanghai Institute of Laser Plasma, China Academy of Engineering Physics, Shanghai 201899, China 
}%

\author{Y-K. Ding}
\affiliation{ 
Institute of Applied Physics and Computational Mathematics, Beijing 100088, China
}%

\author{J. Zheng}%
\affiliation{ 
Department of Plasma Physics and Fusion Engineering and CAS Key Laboratory of Frontier Physics in Controlled Nuclear Fusion, University of Science and Technology of China, Hefei, Anhui 230026, China\looseness=-1
}%
\affiliation{ 
Collaborative Innovation Center of IFSA (CICIFSA), Shanghai Jiao Tong University, Shanghai 200240, China
}%

\date{\today}

\begin{abstract}

Our direct-drive-relevant experiments on the low-coherence Kunwu laser facility identify two-plasmon decay (TPD) as the primary source of hot electrons, and demonstrate for the first time that broadband laser pulses enhance TPD. Using particle-in-cell simulations, we attribute this TPD enhancement and the consequent hot electron production to stochastic intensity spikes inherent in broadband laser fields, robust in both weakly- and strongly-driven regimes. These findings suggest that mitigating hot electron generation requires suppressing these intensity spikes.


\end{abstract}

\maketitle


The achievement of ignition at the National Ignition Facility (NIF) has opened the path toward high-gain inertial confinement fusion (ICF) \cite{Zylstra2022,Abu-shawareb2024} and, ultimately, inertial fusion energy (IFE). Laser-plasma instabilities (LPI) remain a major obstacle, causing laser energy loss \cite{Kritcher2022,Marozas2018} and hot electron preheating, which degrades fuel compression and reduces gain \cite{christopherson2021,christopherson2022}. Broadband lasers are considered a promising route to suppressing LPI \cite{Froula2025}, as early work predicted that LPI growth is inhibited when the laser bandwidth $\Delta \omega \gg \gamma_0$ (the LPI growth rate) \cite{thomson1974,JJThomson1975}, and a few percent bandwidth suffices to significantly reduce LPI \cite{zhao2017,Follett2019,Follett2021}. Broadband laser facilities (Kunwu \cite{Gao2020}, FLUX \cite{Dorrer2020}, PHELIX \cite{Kanstein2025}, and others) have been rapidly developed in recent years\cite{Gao2020b}.
While recent experiments on these facilities with moderate bandwidths ($\sim0.5-0.57\%$) have demonstrated robust suppression of stimulated Brillouin scattering (SBS, a major LPI involving ion acoustic waves) \cite{A.Lei2024PRL,Kanstein2025} and higher energy coupling efficiency \cite{Yang2026}, unexpected hot electron enhancements involving electron plasma waves have also been reported \cite{Wang2024,Kanstein2025}, raising concerns for the application of broadband lasers in ICF.

In direct-drive-relevant experiments performed on the Kunwu facility, the source of hot electrons is still unconfirmed. 
In previous direct-drive experiments on OMEGA and NIF, stimulated Raman scattering (SRS) and two-plasmon decay (TPD)—two major LPIs—have been identified as the primary sources of hot electrons \cite{Rosenberg2018,Michel2013,Hohenberger2015,Rosenberg2023}. Both instabilities generate electron plasma waves that accelerate these hot electrons.
However, previous experiments on the Kunwu facility support neither mechanism.
SRS reflectivities for broadband pulses are consistently at the order of $\sim0.01\%$ \cite{A.Lei2024PRL,Wang2024,Yang2026}, far below typical hot electron energy fraction of $\sim1\%$ at similar intensities\cite{Michel2013}. According to established scaling \cite{Drake1984}, SRS cannot account for such a large hot electron yield.
For TPD, the characteristic $3\omega_0/2$ emission \cite{Seka2009} stays lower for broadband than narrowband lasers \cite{A.Lei2024PRL} , conventionally interpreted as reduced TPD activity, a trend opposite to the observed hot electron enhancement.
This discrepancy motivates the present study to resolve the origin of hot electrons in broadband laser plasma interactions.

In this letter, we identify TPD as the primary source of hot electrons in direct-drive-relevant Kunwu experiments, and demonstrate for the first time that broadband laser pulses enhance TPD. Given the strong angular dependence of the $3\omega_0/2$ emission, which peaks near the polarization plane, we redesign the diagnostic to measure near this plane and observe a clear correlation with hot electron enhancement. Using particle-in-cell (PIC) simulations, we attribute the TPD enhancement and the consequent hot electron production to temporal laser intensity spikes inherent in the broadband laser.

\begin{figure}
\includegraphics[width=1.0\linewidth]{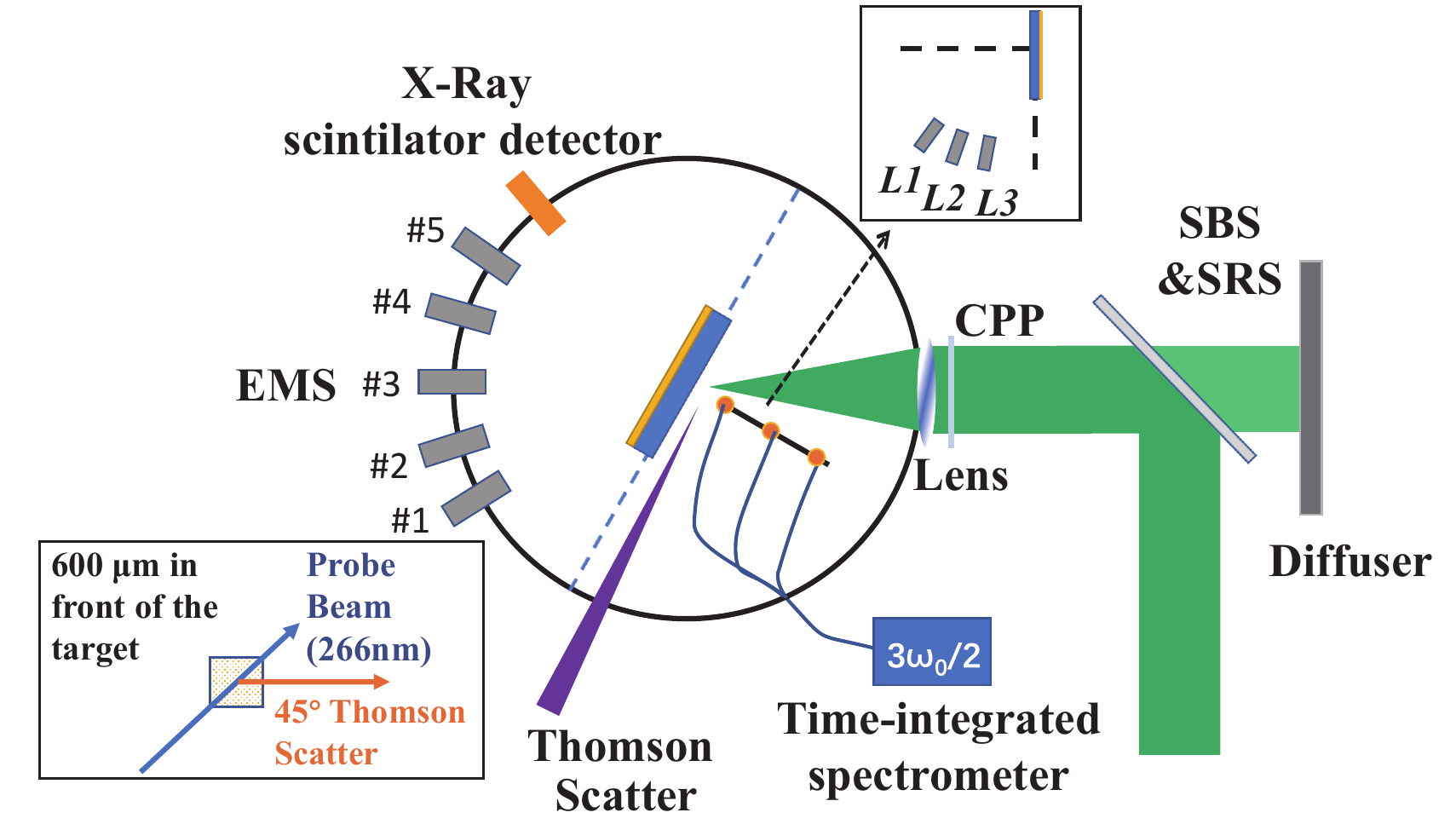}%
\caption{\label{fig:kunwuset} 
Experimental setup. Insets: Thomson scattering optics (lower left) and $3\omega_0/2$ probe layout (upper right).
}
\end{figure}

We perform two rounds of experiments (R2024 and R2025) on the Kunwu platform using CH planar targets with a similar setup, as shown in Fig.~\ref{fig:kunwuset}. Laser pulses with energies of 160--500~J, s-polarization, a flat-top duration of approximately 3~ns irradiate 50-$\mu$m-thick planar parylene-N ($\mathrm{C_8H_8}$) targets backed by 2(R2024)- or 20(R2025)-$\mu$m-thick aluminum layers, at incidence angle $\theta_{in}$ of $30^{\circ}$. 
The central wavelengths are 529~nm (3~nm bandwidth) for broadband (BL) cases and 526~nm (0.01~nm bandwidth) for narrowband (NL) cases\cite{A.Lei2024PRL}.
The laser spot size is  $\sim 220~\mu$m, corresponding to average intensities of $I_0=(1.4-4.4)\times10^{14}~\mathrm{W/cm^2}$. 
\textcolor{black}{To investigate the unexpected hot electron enhancement induced by broadband laser pulses, we use the following diagnostics.}

Hot electron information is diagnosed through two complementary approaches: (i) time-integrated spectra from electron magnetic spectrometers (EMS) located at the rear side of the target, separated by $15^\circ$ from the target center (only EMS $\#$3 is employed for R2024); and (ii) (only for R2025) shot-to-shot relative hot electron energy yields from an X-ray detector measuring hard X-ray bremsstrahlung emission from hot electrons passing through the 20-$\mu$m aluminum layer. The bremsstrahlung photons are filtered by a 300-$\mu$m-thick titanium foil (transmission above $\sim$20~keV) and detected by a BaF$_2$:Y scintillator \cite{Xu2025} coupled to a \textcolor{black}{photomultiplier
, }with time-resolved signals recorded by a fast oscilloscope.


\textcolor{black}{
Since the $3\omega_0/2$ emission is generated by the coupling of the incident laser with TPD-driven plasma waves, it intrinsically exhibits strong angular dependence. In single-beam interactions, TPD dominantly grows and saturates in the laser polarization plane; consequently, the $3\omega_0/2$ emission is also expected to peak near the same plane. Guided by this consideration, we design the diagnostic geometry accordingly.}
The $3\omega_0/2$ emission is collected using multiple probes lying in a common vertical plane through the top and bottom poles of the target chamber, all positioned below the equatorial plane at various angles [upper-right inset of Fig.~\ref{fig:kunwuset}]. The collected light is then diagnosed with a multi-channel spectrometer. The diagnostic plane is perpendicular to the target surface, ensuring that the detection direction is close to the laser polarization plane.


\textcolor{black}{Corona plasma conditions are diagnosed by a Thomson scattering system. A probe beam (energy 100~mJ, wavelength 266~nm, duration $\sim 6$~ns at FWHM, spot size $\sim 50$~$\mu$m) propagates parallel to the target surface at a distance of $\sim 600$~$\mu$m with a spatial accuracy of $\sim$100~$\mu$m. The scattered light is collected at $45^\circ$ to the probe beam, also parallel to the target surface, as shown in the lower-left inset of Fig.~\ref{fig:kunwuset}. The scattered light is then directed to a spectrometer and a streak camera to obtain time-resolved spectra.}

Backscattered SRS and SBS light are directed onto a diffuser plate, collected by fiber probes, and then guided into a spectrometer to obtain time-integrated spectra. 
These spectra are then calibrated to provide the absolute backscattered energy fraction of SRS and SBS.

\begin{figure}
\includegraphics[width=1.0\linewidth]{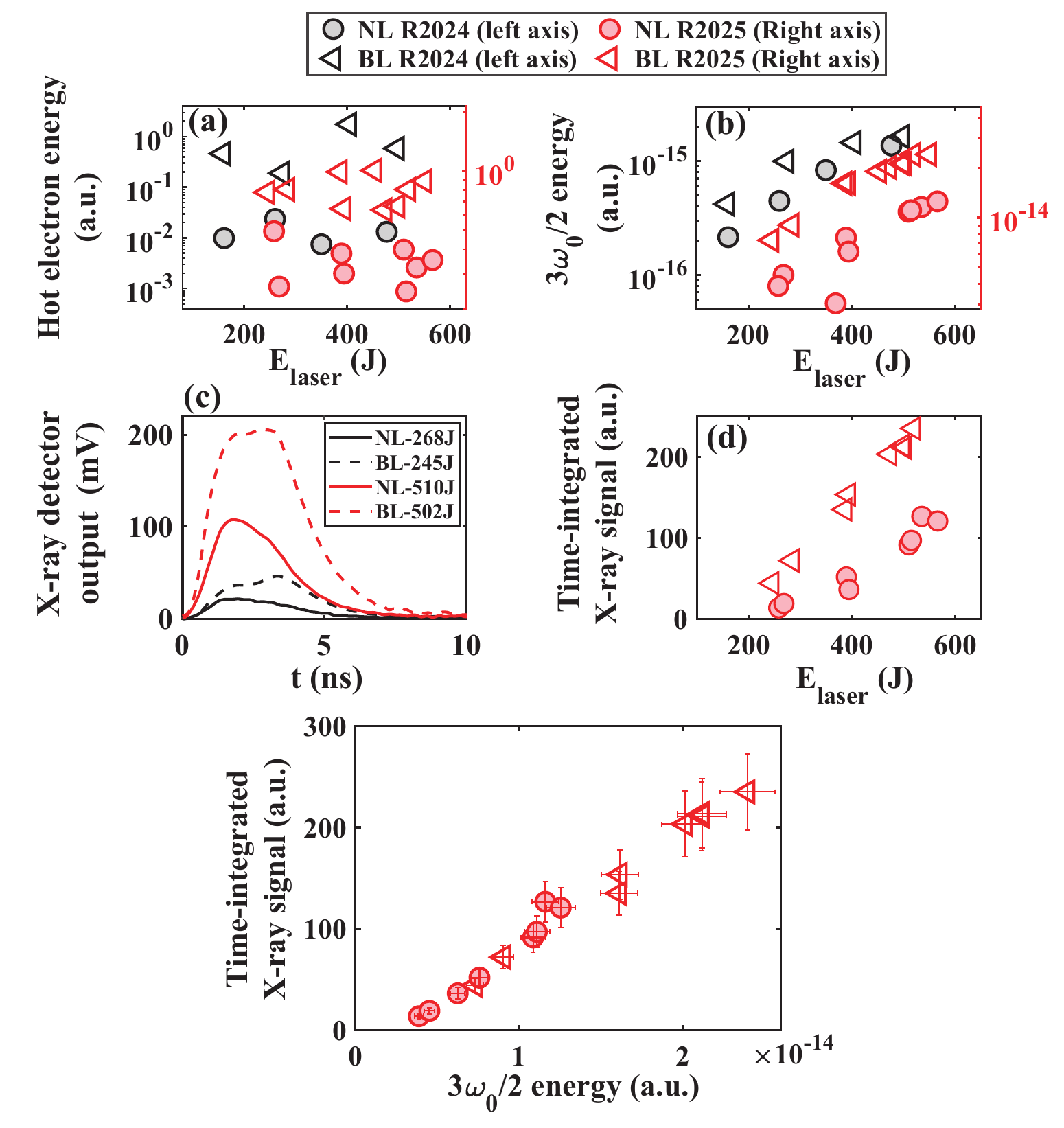}%
\caption{\label{fig:hotE-3_2}
\textcolor{black}{(a) Time-integrated hot electron energy (17\% uncertainty) vs incident laser energy (5\% uncertainty) for all BL and NL shots. Laser energies are marked in the legend.
(b) Time- and spectrally integrated $3\omega_0/2$ scattering energy (7\% uncertainty) vs incident laser energy.
(c) Time-resolved hard X-ray signals for typical NL and BL shots (R2025).
(d) Time-integrated hard X-ray signal (16\% uncertainty) vs incident laser energy.
(e) Time-integrated hard X-ray signal vs the corresponding $3\omega_0/2$ scattering energy from (b). Error bars (7\% and 16\%) are included.
For (a)-(d), error bars are omitted for clarity.}
}
\end{figure}

Figures~\ref{fig:hotE-3_2}(a) and (b) show the total hot electron energy (summed over all EMS channels for R2025, and only EMS $\#$3 for R2024) and the integrated $3\omega_0/2$ scattering energy (summed over L1--L3 probes, time- and spectrally integrated) versus incident laser energy for BL and NL shots, respectively. For NL shots, the hot electron temperature ranges from $30$ to $55$~keV, while for BL shots it increases to $60$ to $120$~keV. 
\textcolor{black}{Across all laser energies, BL irradiation produces markedly higher hot-electron energies and stronger $3\omega_0/2$ scattering than the NL cases.
This trend is independently corroborated by X-ray detector measurements. Fig.~\ref{fig:hotE-3_2}(c) shows typical time-resolved X-ray signals for BL and NL shots with comparable laser energies. The signals rise promptly after the laser arrival and decays within several nanoseconds, limited by the scintillator response. The time-integrated signal can serve as a qualitative indicator for relative hot electron energy.
Fig.~\ref{fig:hotE-3_2}(d) presents the time-integrated X-ray signal versus incident laser energy for all BL and NL shots, confirming that BL consistently yields higher X-ray signals across the entire energy range.}
Fig.~\ref{fig:hotE-3_2}(e) illustrates that the integrated X-ray signal scales monotonically with the integrated $3\omega_0/2$ signal across all shots. 
Remarkably, all data points fall nearly on a tight straight line, revealing a strong relation between TPD activity and hot electron yield. 
All these results evidence TPD as the dominant source of hot electrons in these experiments.
Moreover, the time-integrated backscatter reflectivities range from 1\% to 6\% for SBS and 0.01\% to 0.1\% for SRS \textcolor{black}{(see the End Matter for details)}, suggesting that neither plays a significant role in hot electron generation.


\begin{figure}
\includegraphics[width=0.80\linewidth]{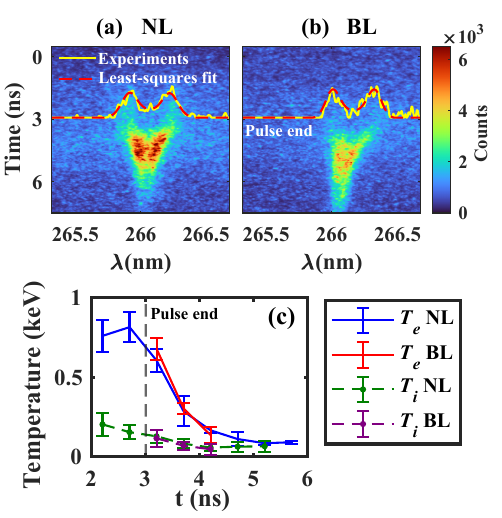}%
\caption{\label{fig:TS} Thomson scattering diagnostics for NL and BL shots. (a) and (b) Time-resolved spectra for the two cases. The yellow solid curves represent spectra integrated over the time window of $3$--$3.5$~ns, with the theoretical fits shown as red dashed curves. (c) Time evolution of $T_e$ and $T_i$ extracted from the spectra.
}
\end{figure}

\textcolor{black}{Thomson scattering data are successfully obtained for two shots} with laser energies of 258~J (NL) and 245~J (BL), as shown in Figs.~\ref{fig:TS}(a) and (b). \textcolor{black}{To our knowledge, this is the first Thomson scattering measurement for plasma states generated by broadband lasers.}
During the heating pulse ($t = 0$--$3$~ns), the NL experiment yields measurable signals from $t = 1$ to $8$~ns, although the signal-to-noise ratio (SNR) is significantly lower during the pulse than after it. For the BL case, the SNR is too low during the heating pulse to extract reliable spectra. For the respective valid time windows, we fit the spectra\cite{Froula2011} to obtain the electron and ion temperatures ($T_e$ and $T_i$) as a function of time.
\textcolor{black}{Fig.~\ref{fig:TS}(c) compares the time evolution of $T_e$ and $T_i$ for the BL and NL cases. During the overlapping time window after the heating pulse ($t > 3$~ns), the two curves agree well, indicating similar plasma states after the pulse ends and suggesting comparable states during the heating pulse as well. These results demonstrate that, under similar plasma conditions, BL lasers drive stronger TPD and higher hot electron yields than NL lasers.}

\textcolor{black}{We investigate the underlying physics using 2D PIC simulations (OSIRIS \cite{Fonseca2002}) with parameters typical of direct-drive ICF. These simulations span both weakly and strongly driven TPD conditions with threshold factors \cite{Simon1983} $\eta\propto I_0 L_n \lambda_0 / T_e$ ranging from 1.3 to 2.6, where $L_n$ and $\lambda_0$ are the density scale length and laser wavelength, respectively. This covers a wide range of TPD behavior, including the regime relevant to our experiments, since TPD-driven hot electron production are primarily governed by $\eta$ \cite{Turnbull2020b}. }

A p-polarized broadband laser with central wavelength $\lambda_0 = 0.351~\mu\mathrm{m}$ is normally incident on a CH plasma with a linear density profile from $0.20n_c$ to $0.27n_c$ over a domain of $35.7~\mu\mathrm{m} \times 45.8~\mu\mathrm{m}$, where $n_c$ is the critical density for the incident laser wavelength. The grid resolution is $\Delta x = \Delta y \approx 0.011 \mu m$, with 200 particles per cell (100 electrons, 50 C$^+$, 50 H$^+$). Periodic boundary conditions are applied in the transverse direction for both fields and particles. In the longitudinal direction (parallel to the laser incident direction), open boundaries are used for fields and thermal boundaries for particles. Hot electrons with energy above 50 keV are diagnosed when crossing the thermal boundaries.

\begin{figure}[tbp]
\includegraphics[width=0.98\linewidth]{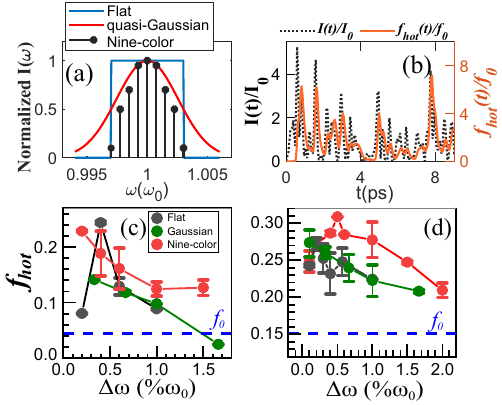}
\caption{\label{fig:diff_I_fhot} 
\textcolor{black}{(a) Flat, quasi-Gaussian, and nine-color laser spectra with $\Delta\omega/\omega_0 = 0.6\%$.}
(b) Instantaneous laser intensity $I(t)$ (black dotted, left axis, normalized by average intensity $I_0$) and hot-electron energy fraction $f_{\mathrm{hot}}(t)$ (orange, right axis, normalized by the narrowband value $f_0$) for the well-above-threshold case with the quasi-Gaussian laser spectrum in (a).
(c) and (d): Simulated $f_{\mathrm{hot}}$ vs bandwidth for near-threshold (c) and well-above-threshold (d) cases. Dots with vertical bars represent mean $f_{\mathrm{hot}}$ and the range from simulations with different random phases. The blue dashed line indicates $f_{\mathrm{hot}}$ for a monochromatic laser at the same average intensity.
}
\end{figure}

For broadband lasers, we approximate continuous spectra using discrete spectral lines. 
We consider three power spectra—flat, quasi-Gaussian, and nine-color—corresponding to configurations on FLUX \cite{Dorrer2020}, Kunwu \cite{Gao2020}, and Shenguang \cite{wangxb2024}, respectively. For the flat and quasi-Gaussian spectra, we set the number of spectral lines to $N_\omega = 100$. \textcolor{black}{Figure~\ref{fig:diff_I_fhot}(a) shows typical broadband spectra with bandwidth $\Delta\omega/\omega_0 = 0.6\%$.}

We consider two plasma conditions sharing $L_n = 150~\mu\mathrm{m}$ and $T_e = 3~\mathrm{keV}$: a strongly driven (well-above-threshold) case with $I_0 = 1.1\times10^{15}~\mathrm{W/cm^2}$, causing the TPD threshold factor $\eta = 2.4$, and a weakly driven (near-threshold) case with $I_0 = 6\times10^{14}~\mathrm{W/cm^2}$ and $\eta = 1.3$. 
For each parameter set, we perform simulations with various $\Delta\omega$ and broadband spectra, focusing on moderate bandwidths $\Delta\omega/\omega_0 \le 2\%$. For each $\Delta\omega$, 2–3 simulations with different initial random phases $\varphi_j$ are run to improve statistical reliability. In all simulations, TPD dominates hot-electron generation, producing hot-electron temperatures of \textcolor{black}{$40$–$90~\mathrm{keV}$}, agreeing with the experiments. 
The backscattered SRS reflectivities range from $0.03\%$ to $0.06\%$, much lower than $f_{\mathrm{hot}}$ [Fig.~\ref{fig:diff_I_fhot}(c-d)], indicating that SRS plays a negligible role in hot electron generation.


Figure~\ref{fig:diff_I_fhot}(b) presents the time evolution of the instantaneous laser intensity $I(t)$ and the resulting hot-electron fraction $f_{\mathrm{hot}}(t)$ for the quasi-Gaussian beam. 
Spectral interference induces pronounced stochastic intensity modulations\cite{Yao2024,Yao2026,Liu2024,Tikhonchuk2026}. Crucially, $f_{\mathrm{hot}}(t)$ is tightly synchronized with these high-intensity spikes, significantly exceeding its values during intensity valleys and the narrowband reference level $f_0$. This synchronization arises because, for a moderate bandwidth ($\Delta\omega/\omega_0 \sim 0.6\%$), the spike duration is comparable to the inverse of the TPD linear growth rate $\gamma_0$ ($\gamma_0 \sim 0.002\omega_0$). This temporal matching allows TPD modes to grow during individual spikes, leading to intermittent but vigorous hot-electron generation that dominates the time-averaged yield. Consequently, the time-averaged hot-electron yield is disproportionately driven by the spikes rather than the average intensity. 

Figures~\ref{fig:diff_I_fhot}(c) and (d) show that, for different broadband schemes and physical parameters, the time-averaged $f_{\mathrm{hot}}$'s are substantially higher than in the single-frequency case. For the near-threshold case, this holds for $\Delta\omega/\omega_0 \le 1.7\%$, the upper bound of our moderate-bandwidth range, qualitatively agreeing with Kunwu experiments using a quasi-Gaussian spectrum with $\Delta\omega/\omega_0 = 0.57\%$ \cite{Wang2024}. As $\Delta\omega$ increases further, $f_{\mathrm{hot}}$ drops below $f_0$ because larger bandwidth progressively impedes the three-wave matching for TPD. For the well-above-threshold case, $f_{\mathrm{hot}} > f_0$ persists even at $\Delta\omega/\omega_0 = 2\%$.
Figures~\ref{fig:diff_I_fhot}(c) and (d) also reveals a trade-off for controlling TPD under weakly driven conditions: while $f_{\mathrm{hot}}$ is more readily suppressed below $f_0$ at sufficiently high $\Delta\omega$, an insufficient $\Delta\omega$ can amplify it far above $f_0$. 
The reason is that, the hot-electron energy generated by the spikes is largely insensitive to whether the average intensity $I_0$ approaches threshold, since the spikes themselves far exceed $I_0$. In contrast, for a narrowband laser, $f_0$ drops sharply with $I_0$ in the near-threshold regime\cite{Michel2013}.

\textcolor{black}{Since intensity spikes dominate hot electron production, mitigating hot electron generation requires suppressing these spikes. This can be achieved by tailoring their characteristics: reducing the spike width requires increasing the laser bandwidth\cite{Ai2026}, while reducing the spike height calls for dedicated methods and techniques.}


In summary, we have identified TPD as the primary source of hot electrons in direct-drive-relevant Kunwu experiments, and demonstrated for the first time that broadband laser pulses enhance TPD. Using PIC simulations, we attribute this TPD enhancement and the resulting hot electron production to intensity spikes inherent in broadband laser fields. These findings provide a physical basis for developing strategies to suppress hot electrons by controlling intensity spike characteristics, offering practical insights for the design of future ICF laser systems.

We gratefully acknowledge the beneficial assistance and help of all the technical staff during the experiments. 
\textcolor{black}{We thank Prof. X. H. Yuan for the support on the experiments.}
This work was supported by the National Natural Science Foundation of China (NSFC) under Grant Nos. 12275269, 12388101, 12375243, U2430207 and 12275032, Science Challenge Project under Grant No. TZ2025014, National Key R\&D Program of China, No. 2023YFA1608400 and Strategic Priority Research Program of the Chinese Academy of Sciences under Grant Nos. XDA25010200 and XDA25050400. The numerical simulations were performed on the Hefei Advanced Computing Center. We thank the UCLA-IST OSIRIS Consortium for the use of OSIRIS.

\bibliography{apssamp}

  \vspace{1cm}
  \onecolumngrid
  \begin{center}
  \textbf{\large End Matter}
 \end{center}

  \twocolumngrid
  


\begin{figure}[h]
      \centering
\includegraphics[width=0.95\linewidth]{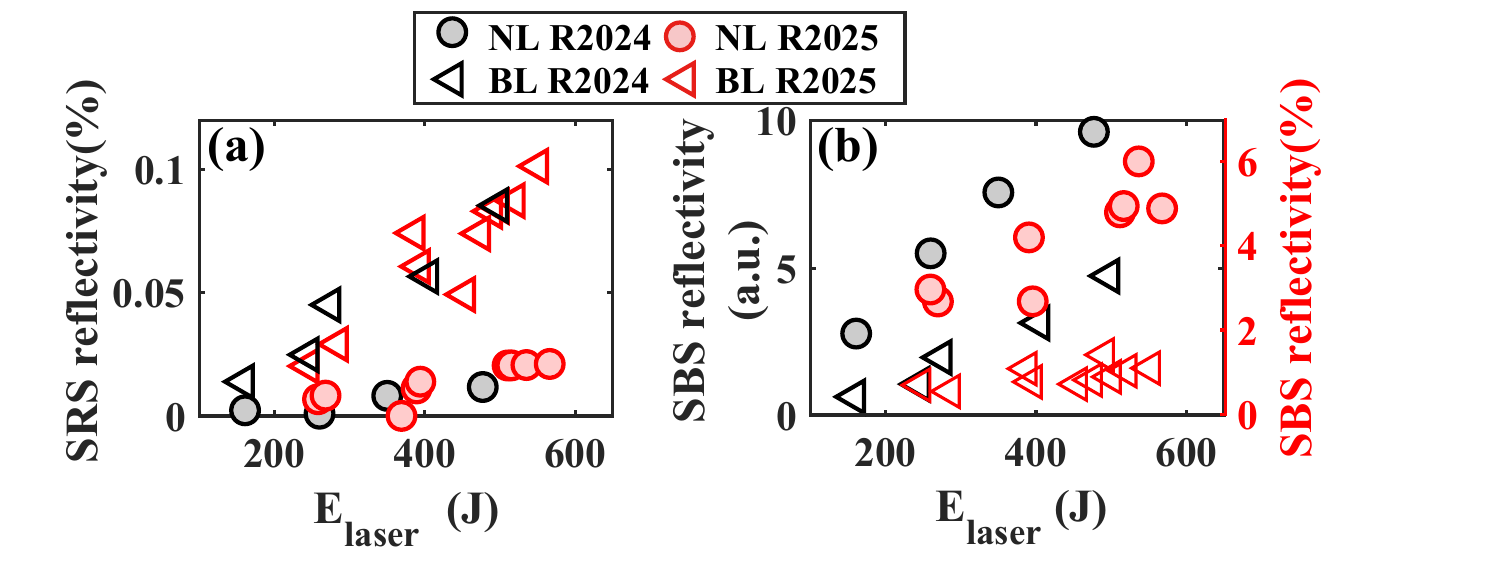}
      \caption{Time-integrated backscattered (a) SRS and (b) SBS reflectivities versus laser energy for R2024 and R2025. In (b), R2024 SBS data (black markers) are in arbitrary units.}
      \label{fig:placeholder}
  \end{figure}

The time-integrated backscattered SRS and SBS reflectivities for R2024 and R2025 are plotted in Fig.~\ref{fig:placeholder} as a function of laser energy. Absolute values are available for R2025 (both SRS and SBS) and for R2024 SRS; R2024 SBS data (black circles and triangles in Fig.~\ref{fig:placeholder}b) are presented in arbitrary units. 
The results show consistent SRS enhancement and SBS suppression for BL cases compared to NL cases, with SRS reflectivities reaching up to $\sim0.1\%$ for BL and SBS up to $\sim6\%$ for NL.

  
\end{document}